# An Improved Self-Consistent One-Dimensional Slender Jet Model

**Leonid Pekker**[a)]

FujiFilm Dimatix Inc., Lebanon NH 03766, USA

**David Pekker**

University of Pittsburgh, Pittsburgh PA 15260, USA

[a)]Author to whom correspondence should be addressed: leonid.pekker@fujifilm.com

**Abstract**

In 1994, Eggers and Dupont suggested the slender jet model, a one-dimensional model that describes the motion of a thin axisymmetric column of viscous, incompressible fluid with a free surface. In their model, the momentum equation was derived in a manner that was not completely self-consistent. Consequently, the viscosity term was described with less accuracy than the surface tension term. In this paper, we derive a novel slender jet momentum equation in a completely self-consistent manner that allows us to describe both the viscosity and the surface tension forces with the same accuracy as the surface tension term in the Eggers-Dupont model. Our derivation does not affect the volume conservation equation, which remains identical to the one in the Eggers-Dupont model. We show that our model predicts different Plateau-Rayleigh instability dynamics as compared to the Eggers-Dupont's model. The differences between the models are particularly large at small Reynolds numbers, where the viscosity plays a prominent role in development the Plateau-Rayleigh instability.

# 1. Introduction

In their pioneering work [1], Eggers and Dupont derived a set of slender jet equations describing a thin axisymmetric column of incompressible liquid with a free surface in a lubrication approximation. This model has been widely used in numerical simulations of drop formations and compared with 2D – CFD simulations [2, 3,7, 9, 12] and experiments [1, 4-6, 8, 11].

In [1], $h(z,t)$, the jet radius, and $v_z(z,t)$, the liquid longitudinal velocity averaged over the cross-section of the jet, are calculated from the momentum and volume conservation slender jet partial differential equations; $z$ is the axis directed along the jet; and $t$ is time, Fig. 1. As discussed in [1], at the tip of a jet and near a neck (i.e., at a point where a droplet is breaking off) $|\partial h/\partial z| \to \infty$ and $h \to 0$, see Fig.1; this leads to singularities in the derivation of the slender jet momentum equation [1]. The authors took this singularity into account in a consistent way while they obtained the surface tension term, but not in deriving the viscosity term. Therefore, in their model, the surface tension forces are described with greater accuracy than the viscosity forces.

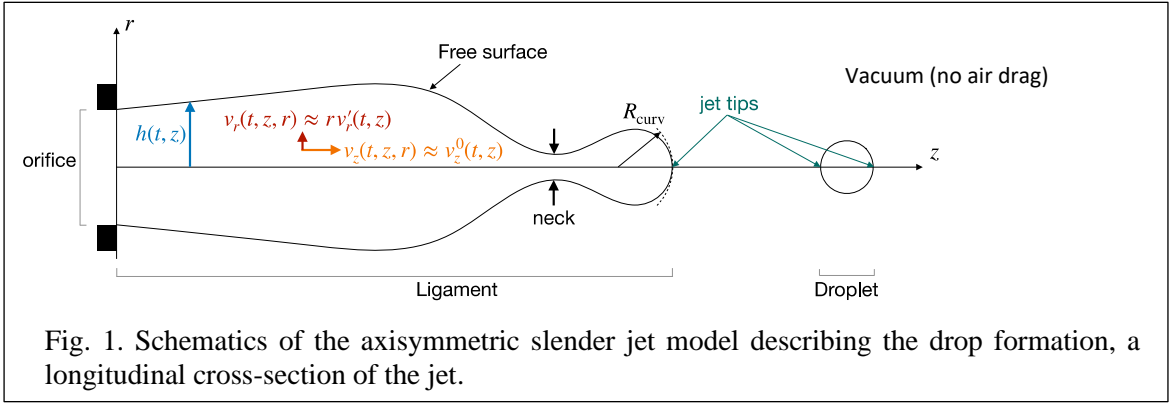

Fig. 1. Schematics of the axisymmetric slender jet model describing the drop formation, a longitudinal cross-section of the jet.

In our paper, we derive the slender jet momentum equation in a self-consistent manner that allows us to obtain both surface tension and viscosity terms with the same accuracy as the surface tension term in [1]. The volume conservation equations in both models are the same.

In Section II, we derive a new set of slender jet equations. We show that far from a tip or a neck of the jet, where $\partial h/\partial z \ll 1$, both sets of slender jet equations are the same. The difference between the two models becomes significant near tips and necks of the jet, where $\partial h/\partial z$ becomes large. In Section III, we numerically simulate the Plateau-Rayleigh instability of a liquid axisymmetric column using both slender jet models. The conclusions are presented in Section IV.

## II. Slender Jet Equations

As in [1], we start from the incompressible Navier-Stokes and continuity equations for an axisymmetric liquid column expressed in cylindrical coordinates,

$$\frac{\partial v_r}{\partial t} + v_r \frac{\partial v_r}{\partial r} + v_z \frac{\partial v_r}{\partial z} = -\frac{1}{\rho} \cdot \frac{\partial P}{\partial r} + \frac{\mu}{\rho} \left( \frac{1}{r}\frac{\partial}{\partial r}\left(r\frac{\partial v_r}{\partial r}\right) - \frac{v_r}{r^2} + \frac{\partial^2 v_r}{\partial z^2} \right), \quad (1)$$

$$\frac{\partial v_z}{\partial t} + v_r \frac{\partial v_z}{\partial r} + v_z \frac{\partial v_z}{\partial z} = -\frac{1}{\rho} \cdot \frac{\partial P}{\partial z} + \frac{\mu}{\rho} \left( \frac{1}{r}\frac{\partial}{\partial r}\left(r\frac{\partial v_z}{\partial r}\right) + \frac{\partial^2 v_z}{\partial z^2} \right), \quad (2)$$

$$\frac{\partial v_z}{\partial z} + \frac{1}{r}\frac{\partial}{\partial r}(rv_r) = 0, \quad (3)$$

where $v_r = v_r(t,z,r)$ is the velocity in the radial direction, $v_z = v_z(t,z,r)$ is the velocity along the axis, $P = P(t,z,r)$ is the pressure, $\rho$ is the density, and $\mu$ is the viscosity; and $v_r$ at the axis is equal to zero,

$$v_r(r=0) = 0 \quad (4)$$

The free surface boundary conditions for Eq. (1) – (3) are

$$\vec{n}\overleftrightarrow{S}\vec{\tau} = 0, \quad (5)$$

$$P_h - \vec{n}\overleftrightarrow{S}\vec{n} = \gamma\left(\frac{1}{R_1} + \frac{1}{R_2}\right) \quad (6)$$

$$\frac{\partial h}{\partial t} + \frac{\partial h}{\partial z}(v_z)_h = (v_r)_h, \quad (7)$$

where $\vec{n}$ and $\vec{\tau}$ are the normal and tangential vectors to the surface respectively; $\overleftrightarrow{S}$ is the stress tensor at the surface; $R_1$ and $R_2$ are the surface axial and transverse curvature radii; $\gamma$ is the surface tension; $h$ is the radius of the jet; and index $h$ corresponds to the surface such that $(v_z)_h$ and $(v_r)_h$ are the liquid velocities at the jet surface and $P_h$ is the liquid pressure at the surface. Eq. (5) states that the tangential stress at the surface is equal to zero; Eq. (6) states that the normal stress at the surface is balanced by surface tension and pressure; and Eq. (7) states that the surface is moving with the same velocity as the liquid at the surface. Eqs. (5) and (6) can be presented in the following form [1]

$$2\frac{\partial h}{\partial z}\left(\frac{\partial v_r}{\partial r} - \frac{\partial v_z}{\partial z}\right)_h + \left(1 - \left(\frac{\partial h}{\partial z}\right)^2\right)\left(\frac{\partial v_r}{\partial z} + \frac{\partial v_z}{\partial r}\right)_h = 0, \quad (8)$$

$$P_h - \left(\frac{2\mu}{\left(\frac{\partial h}{\partial z}\right)^2+1}\right)\left(\left(\frac{\partial v_r}{\partial r}\right)_h - \left(\frac{\partial v_z}{\partial r}+\frac{\partial v_r}{\partial z}\right)_h \frac{\partial h}{\partial z} + \left(\frac{\partial v_z}{\partial z}\right)_h \left(\frac{\partial h}{\partial z}\right)^2\right) = \gamma \frac{1+\left(\frac{\partial h}{\partial z}\right)^2 - h\frac{\partial^2 h}{\partial z^2}}{h\left(1+\left(\frac{\partial h}{\partial z}\right)^2\right)^{1.5}}. \qquad (9)$$

In the right hands side of Eq. (9), we have used an explicit expression for the curvature of the jet [1]. Thus, the set of Eqs. (1) – (4) along with free surface boundary conditions, Eqs. (7) – (9), describes the axisymmetric jet with a free surface.

The next step, in our method of deriving a slender jet momentum equation, is to integrate Eq. (2) over the cross-section of the jet,

$$\int_0^h \left(\frac{\partial v_z}{\partial t} + v_r \frac{\partial v_z}{\partial r} + v_z \frac{\partial v_z}{\partial z}\right) r dr = \int_0^h \left(-\frac{1}{\rho}\cdot\frac{\partial P}{\partial z} + \frac{\mu}{\rho}\left(\frac{1}{r}\frac{\partial}{\partial r}\left(r\frac{\partial v_z}{\partial r}\right) + \frac{\partial^2 v_z}{\partial z^2}\right)\right) r dr. \qquad (10)$$

As shown in Appendix A, using Eq. (3) along with the boundary conditions at the liquid surface, Eqs. (7) – (9), and the fact that

$$\frac{\partial}{\partial t}\int_0^{h(t,z)} F(t,z,x')dx' = \int_0^{h(t,z)} \frac{\partial F(t,z,x')}{\partial t} dx' + F(t,z,h(z,t))\frac{\partial h(z,t)}{\partial t},$$

Eq. (10) can be transformed into divergence form:

$$\frac{\partial}{\partial t}\left(\int_0^h v_z r dr\right) + \frac{\partial}{\partial z}\left(\int_0^h r\left(v_z^2 + \frac{P}{\rho}\right) dr - \frac{\gamma}{\rho}\frac{h}{\left(1+\left(\frac{\partial h}{\partial z}\right)^2\right)^{1/2}} + \frac{2\mu}{\rho} h(v_r)_h\right) = 0. \qquad (11)$$

Now let us consider a first order slender jet model defined by the following approximation

$$P(t,z,r) = P_0(t,z), \quad v_r(t,z,r) = rv_r'(t,z), \quad v_z(t,z,r) = v_z^0(t,z). \qquad (12)$$

That is the $P$ and $v_z$ are assumed to be independent of the radial coordinate $r$, while $v_r$ is assumed to have linear dependence on $r$, which is required to satisfy Eq. (4). This approximation is valid when the radial velocities in the jet are smaller than the longitudinal velocities:

$$v_r \ll v_z. \qquad (13)$$

Substituting $v_r$ and $v_z$ from Eq. (12) into Eq. (3), we obtain

$$v_r = -\frac{r}{2}\frac{\partial v_z^0}{\partial z}. \qquad (14)$$

Substituting $v_r$ from Eq. (14) and $v_z$ from Eq. (12) into Eq. (9) and taking into account, from Eq. (12), that $P_h = P_0$, we obtain the following equation for $P_0$:

$$P_0 = \left(\frac{2\mu}{\left(\frac{\partial h}{\partial z}\right)^2+1}\right)\left(-\frac{1}{2}\frac{\partial v_z^0}{\partial z} + \frac{1}{2}\frac{\partial^2 v_z^0}{\partial z^2}h\frac{\partial h}{\partial z} + \frac{\partial v_z^0}{\partial z}\left(\frac{\partial h}{\partial z}\right)^2\right) + \gamma\frac{1+\left(\frac{\partial h}{\partial z}\right)^2 - h\frac{\partial^2 h}{\partial z^2}}{h\left(1+\left(\frac{\partial h}{\partial z}\right)^2\right)^{1.5}}. \tag{15}$$

Noting that at the tip of a jet where $h \to 0$ and $|\partial h/\partial z| \to \infty$, Eq. (15) has several numerical issues. Specifically, both the viscosity term and the surface tension term have numerators and denominators that diverge as $\left|\frac{\partial h}{\partial z}\right|^2$. Additionally, the viscosity term contains the numerically ill-defined quantity $\frac{1}{2}\frac{\partial^2 v_z^0}{\partial z^2}h\frac{\partial h}{\partial z}$. While it is possible to analyze these issues one a time to make them numerically well-defined, we found a more elegant approach is to introduce a new variable

$$y = h^2. \tag{16}$$

Substituting Eq. (16) into Eq. (15), we obtain the following equation for $P_0$,

$$P_0 = \left(\frac{2\mu}{y+\frac{1}{4}\left(\frac{\partial y}{\partial z}\right)^2}\right)\left(-\frac{y}{2}\frac{\partial v_z^0}{\partial z} + \frac{1}{4}\frac{\partial^2 v_z^0}{\partial z^2}y\frac{\partial y}{\partial z} + \frac{1}{4}\frac{\partial v_z^0}{\partial z}\left(\frac{\partial y}{\partial z}\right)^2\right) + \gamma\frac{y+\frac{1}{2}\left(\frac{\partial y}{\partial z}\right)^2 - \frac{y}{2}\frac{\partial^2 y}{\partial z^2}}{\left(y+\frac{1}{4}\left(\frac{\partial y}{\partial z}\right)^2\right)^{1.5}}. \tag{17}$$

Near the tip of a jet, the surface is described by $h = \sqrt{2R_{\text{curv}}(z_{\text{tip}} - z)}$ or $y = 2R_{\text{curv}}(z_{\text{tip}} - z)$. Therefore, at the tip

$$y_{\text{tip}} = 0, \qquad \left(\frac{\partial y}{\partial z}\right)_{\text{tip}} = -2R_{\text{curv}}, \qquad \left(\frac{\partial^2 y}{\partial z^2}\right)_{\text{tip}} = -2, \tag{18}$$

and hence we observe that all the numerical issues of Eq. (15) have disappeared in Eq. (17).

To obtain the slender jet momentum equation, we substitute $v_z$ from Eq. (12), $v_r$ from Eq. (14), into Eq. (11), and set $P = P_0$ from Eq. (17), to obtain

$$\frac{\partial}{\partial t}(v_z^0 y) + \frac{\partial}{\partial z}\left(y(v_z^0)^2 - \frac{\gamma}{\rho}y^2\frac{1+\frac{1}{2}\frac{\partial^2 y}{\partial z^2}}{\left(y+\frac{1}{4}\left(\frac{\partial y}{\partial z}\right)^2\right)^{1.5}} - \frac{\mu}{\rho}\frac{y^2}{y+\frac{1}{4}\left(\frac{\partial y}{\partial z}\right)^2}\left(3\frac{\partial v_z^0}{\partial z} - \frac{1}{2}\frac{\partial y}{\partial z}\frac{\partial^2 v_z^0}{\partial z^2}\right)\right) = 0. \tag{19}$$

Substituting $v_z$ from Eq. (12) and $v_r$ from Eq. (14) into Eq. (7), we obtain the volume conservation equation [1]. Multiplying this equation by $h$, we obtain the volume conservation equation [1] in a divergent form

$$\frac{\partial y}{\partial t} + \frac{\partial}{\partial z}(yv_z^0) = 0, \tag{20}$$

Thus, Eqs. (19) and (20) are the new set of slender jet equations.

Dropping the second term in both the numerator and denominator of the viscosity term of Eq. (19), we obtain the following slender jet equation,

$$\frac{\partial}{\partial t}(v_z^0 y) + \frac{\partial}{\partial z}\left(y(v_z^0)^2 - \frac{\gamma}{\rho} y^2 \frac{1+\frac{1}{2}\frac{\partial^2 y}{\partial z^2}}{\left(y+\frac{1}{4}\left(\frac{\partial y}{\partial z}\right)^2\right)^{1.5}} - \frac{3\mu}{\rho} y \frac{\partial v_z^0}{\partial z}\right) = 0, \quad (21)$$

which, as shown in Appendix B, is equivalent to the momentum equation [1], but written in a divergent form.

For numerical simulations, it is convenient to introduce a new variable

$$M = v_z^0 y, \quad (22)$$

which is the liquid momentum in $z$-direction integrated over the cross-section of the jet multiplied by two and divided by liquid density and $\pi$. Substituting $v_z^0 = M/y$ into Eqs. (19), (20) leads to the following set of slender jet equations:

$$\frac{\partial M}{\partial t} + \frac{\partial}{\partial z}\left(\frac{M^2}{y} - \frac{\gamma y^2}{\rho}\frac{1+\frac{1}{2}\frac{\partial^2 y}{\partial z^2}}{\left(y+\frac{1}{4}\left(\frac{\partial y}{\partial z}\right)^2\right)^{1.5}} - \frac{\mu}{\rho}\frac{y}{y+\frac{1}{4}\left(\frac{\partial y}{\partial z}\right)^2}\right.$$

$$\left.\left(3\left(\frac{\partial M}{\partial z} - \frac{M}{y}\frac{\partial y}{\partial z}\right) - \frac{1}{2}\frac{\partial y}{\partial z}\left(\frac{\partial^2 M}{\partial z^2} - \frac{2}{y}\frac{\partial M}{\partial z}\frac{\partial y}{\partial z} + \frac{2M}{y^2}\left(\frac{\partial y}{\partial z}\right)^2 - \frac{M}{y}\frac{\partial^2 y}{\partial y^2}\right)\right)\right) = 0. \quad (23)$$

$$\frac{\partial y}{\partial t} + \frac{\partial M}{\partial z} = 0. \quad (24)$$

In new variables, the Eggers-Dupont momentum Eq. (21) reduces to the form

$$\frac{\partial M}{\partial t} + \frac{\partial}{\partial z}\left(\frac{M^2}{y} - \frac{\gamma y^2}{\rho}\frac{1+\frac{1}{2}\frac{\partial^2 y}{\partial z^2}}{\left(y+\frac{1}{4}\left(\frac{\partial y}{\partial z}\right)^2\right)^{1.5}} + \frac{3\mu}{\rho}\left(\frac{\partial M}{\partial z} - \frac{M}{y}\frac{\partial y}{\partial z}\right)\right) = 0 \quad (25)$$

Now let consider the boundary conditions at the tip of a jet (or a droplet). At the tip, the boundary conditions for Eqs. (23) and (24) are trivial:

$$y_{\text{tip}} = 0 \text{ and } M_{\text{tip}} = 0, \quad (26)$$

These boundary conditions must be supplemented by an equation describing how the position of the tip $z_{\text{tip}}(t)$ changes in time

$$\frac{\partial z_{\text{tip}}}{\partial t} = v_{\text{tip}}. \tag{27}$$

Here, $v_{\text{tip}}$ is the velocity of the tip. Substituting Eq. (22) into $(\partial M/\partial z)_{\text{tip}}$ and using that $y_{\text{tip}} = 0$, we obtain an equation for $v_{\text{tip}}$

$$v_{\text{tip}} = \frac{\left(\frac{\partial M}{\partial z}\right)_{\text{tip}}}{\left(\frac{\partial y}{\partial z}\right)_{\text{tip}}}. \tag{28}$$

We conclude our discussion of boundary conditions by noting that the boundary conditions at the point where a jet comes out of an orifice, e.g. on the left hand side of the ligament depicted in Fig. 1, are

$$y_0(t) = h_0^2(t), \quad M_0 = h_0^2(t)v_0(t). \tag{29}$$

## III. Numerical Results

To illustrate the differences between our slender jet model and the Eggers-Dupont model [1], we consider the case of breakup of an infinite column with an initial sinusoidal perturbation on the radius of the column,

$$h = h_0\left(1 + \varepsilon \cos\left(\frac{2\pi z}{\lambda}\right)\right). \tag{30}$$

As in [7,12,14], we switch to dimensionless variables associated with the jet

$$\tilde{y} = \frac{y}{h_0^2}, \quad \tilde{h} = \frac{h}{h_0}, \quad \tilde{t} = \frac{t}{(\rho h_0^3/\gamma)^{1/2}}, \quad \tilde{z} = \frac{z}{h_0}, \quad \widetilde{M} = M\left(\frac{\rho}{\gamma h_0^3}\right)^{1/2}, \quad \tilde{v} = v\left(\frac{\gamma}{\rho h_0}\right)^{1/2}, \quad \text{ReN} = \frac{\sqrt{\rho h_0 \gamma}}{\mu}. \tag{31}$$

Dropping the "~" from the new variables, we express equations (23) – (25) in dimensionless form

$$\frac{\partial M}{\partial t} + \frac{\partial}{\partial z}\left(\frac{M^2}{y} - y^2 \frac{1+\frac{1}{2}\frac{\partial^2 y}{\partial z^2}}{\left(y+\frac{1}{4}\left(\frac{\partial y}{\partial z}\right)^2\right)^{1.5}} - \frac{1}{\text{ReN}}\frac{y}{y+\frac{1}{4}\left(\frac{\partial y}{\partial z}\right)^2}\right.$$

$$\left.\left(3\left(\frac{\partial M}{\partial z} - \frac{M}{y}\frac{\partial y}{\partial z}\right) - \frac{1}{2}\frac{\partial y}{\partial z}\left(\frac{\partial^2 M}{\partial z^2} - \frac{2}{y}\frac{\partial M}{\partial z}\frac{\partial y}{\partial z} + \frac{2M}{y^2}\left(\frac{\partial y}{\partial z}\right)^2 - \frac{M}{y}\frac{\partial^2 y}{\partial y^2}\right)\right)\right) = 0, \tag{32}$$

$$\frac{\partial y}{\partial t} + \frac{\partial M}{\partial z} = 0, \tag{33}$$

$$\frac{\partial M}{\partial t} + \frac{\partial}{\partial z}\left(\frac{M^2}{y} - y^2 \frac{1+\frac{1}{2}\frac{\partial^2 y}{\partial z^2}}{\left(y+\frac{1}{4}\left(\frac{\partial y}{\partial z}\right)^2\right)^{1.5}} - \frac{3}{\text{ReN}}\left(\frac{\partial M}{\partial z} - \frac{M}{y}\frac{\partial y}{\partial z}\right)\right) = 0. \qquad (34)$$

In Eqs. (32) and (34), ReN plays the role of the Reynolds number, which features in the breakup of the jet caused by the Plateau-Rayleigh instability. In defining ReN, $\sqrt{\gamma/\rho h_0}$ and $h_0$ play the roles of the characteristic velocity and the characteristic length, respectively.

We consider two cases: (a) ReN = 100 which corresponds to the case of a low-viscosity liquids and (b) ReN = 0.1, the case of a high-viscosity liquids. In both cases, the dimensionless wave number $k = 2\pi h_0/\lambda$ is chosen as 0.7, which corresponds to the maximum rate of the Plateau-Rayleigh instability for inviscid liquid column in linear approximation [13]; $\varepsilon$ is chosen as 0.05. The results of the simulation are presented in Figs. 2 and 3. As expected, in the case of the large ReN (large Reynold number), the differences between the Eggers-Dupont's slender jet model and ours are very small, Fig. 2, because the viscosity plays a minor role in the development of the Plateau-Rayleigh instability. Only at the breakup point, at $t = 9.65$ in Fig. 2, do the velocities differ enough to be noticed; this is not surprising because of the large gradients in the radius of the jet at its neck-bottle where the viscosity forces are very large. In the case of the small $ReN$ (small Reynolds number), as shown in Fig. 3, at the initial stage of the development of the Plateau-Rayleigh instability, the differences between the models are small because the lubrication approximation, $\partial h/\partial z \ll 1$, is well satisfied at those times, and Eqs. (32) and (34) are the identical. However, at the final stage of the breakup of the liquid column where $\partial h/\partial z$ becomes large, the differences between our model and the Eggers-Dupont's slender jet model become significant, Fig. 3. Also, as seen in Fig. 3, the rate of development of the Plateau-Rayleigh instability calculated by our model is larger than that calculated by model [1] and, correspondingly, the breakup time of the jet calculated by our model is 305 and by Eggers-Dupont's model is 312.

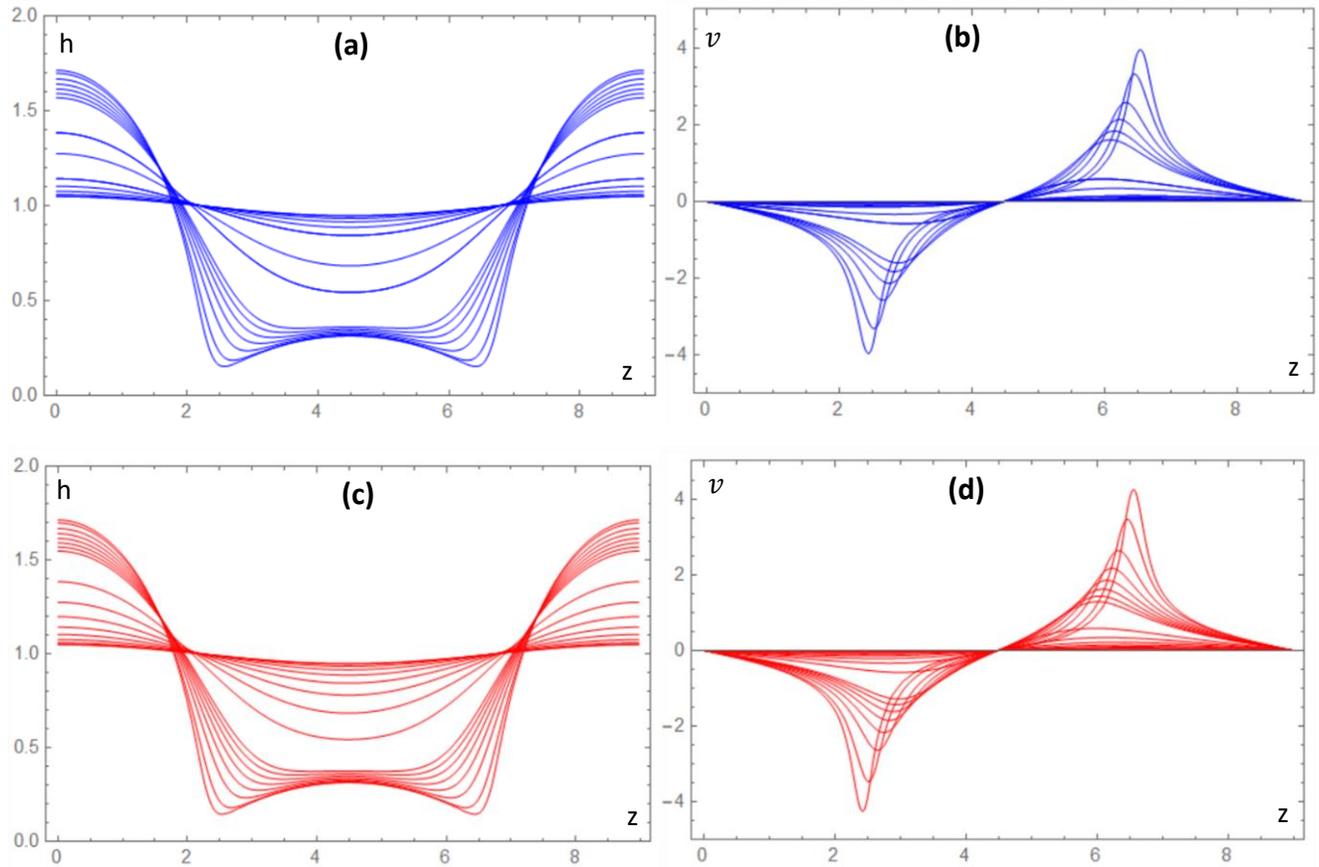

Fig. 2. The dynamics of the breakup of an infinite liquid column with an initial sinusoidal perturbation: $Ren = 100$, $k = 0.7$, $\varepsilon = 0.05$. Panels a and b: free surface shape and axial velocity obtained from the Eggers-Dupont model, Eqs. (33) and (34). Panels c and d: free surface shape and velocity obtained from our model, Eqs. (32) and (33). The lines plotted corresponds to times t = 0, 1, 2, 3, 4, 5, 6, 7, 8, 9, 9.1. 9.2, 9.3. 9.4, 9.5, 9.6, 9.65. We note that if we superimpose panel a and c then all the lines would coincide. If we overlap panels b and d all the lines would coincide except the ones corresponding to t = 9.65 which would slightly deviate from each other.

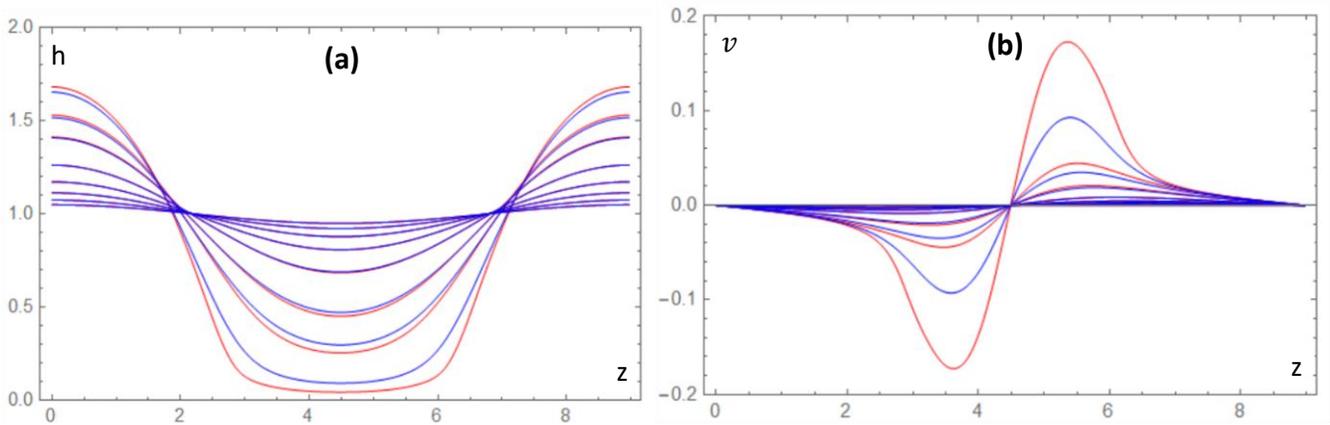

Fig. 3. The Dynamics of breakup of a liquid infinite column with sinusoidal perturbation, Eq. 34: $Ren = 0.1$, $k = 0.7$, $\varepsilon = 0.05$; blue plots: Eggers-Dupont model and red plots: our model. The line corresponds to time = 0, 50, 100, 150, 200, 250, 275, 300. Panel a - $h(z, t)$ and Panel b - $v(z, t)$.

## IV. Conclusions

In this paper, we derived the set of slender jet equations in a self-consistent manner which differs from the Eggers-Dupont's model [1]. Unlike in [1], in our model, both the surface tension and the viscosity terms in the jet momentum equation are obtained with the same accuracy. To illustrate the differences between the models, we have simulated the breakup process of an infinite liquid column with an initial sinusoidal perturbation. We showed that in the early stages of the breakup, while $|\partial h/\partial z| \ll 1$, both models agree. However, in the late stages of breakup the models predictions deviate. The deviations are especially large when viscosity dominates, i.e., for $\text{ReN} \lesssim 1$.

**Acknowledgements**

The authors would like to express their sincere gratitude to Dan Barnett, Paul Hoisington, James Myrick, and Matthew Aubrey for helpful discussions. Special thanks to Chris Menzel who initiated this research.

**Data Availability**

The data that support the findings of this study are available from the corresponding author upon reasonable request.

**Appendix A: Derivation Eq. (11)**

Let us present the $LHS$ of Eq. (10) as $LHS = A + B$, where

$$A = \int_0^h \frac{\partial v_z}{\partial t} r\, dr = \frac{\partial}{\partial t}\left(\int_0^h v_z\, r\, dr\right) - v_z h \frac{\partial h}{\partial t} \tag{A1}$$

$$B = \int_0^h \left(v_r \frac{\partial v_z}{\partial r} + v_z \frac{\partial v_z}{\partial z}\right) r\, dr = \int_0^h \left(\frac{\partial(v_z v_r r)}{\partial r} - v_z \frac{\partial(v_r r)}{\partial r} + r v_z \frac{\partial v_z}{\partial z}\right) dr =$$

$$= \int_0^h \left(\frac{\partial(v_z v_r r)}{\partial r} + r v_z \frac{\partial v_z}{\partial z} + r v_z \frac{\partial v_z}{\partial z}\right) dr = \int_0^h \left(\frac{\partial(v_z v_r r)}{\partial r} + \frac{\partial(r v_z^2)}{\partial z}\right) dr =$$

$$= (v_z v_r)_h h + \frac{\partial}{\partial z}\int_0^h v_z^2 r\, dr - h \frac{\partial h}{\partial z}(v_z^2)_h, \tag{A2}$$

where index $h$ corresponds to the jet surface that

$$\big(F(t,r,z)\big)_h = F(t,h(t,z),z). \tag{A3}$$

In Eqs. (A1) and (A2), we have used Eq. (3) and the following formula

$$\frac{\partial}{\partial t}\int_0^{h(t,z)} F(t,z,x')dx' = \int_0^{h(t,z)} \frac{\partial F(t,z,x')}{\partial t}dx' + F(t,z,h(z,t))\frac{\partial h(z,t)}{\partial t}. \tag{A4}$$

Collecting all terms of Eqs. (A1) and (A2) we obtain

$$LHS = A + B = \frac{\partial}{\partial t}\left(\int_0^h v_z\, rdr\right) + \frac{\partial}{\partial z}\int_0^R v_z^2 rdr - (v_z)_h R\left(\frac{\partial h}{\partial t} + (v_z)_h\frac{\partial h}{\partial z} - (v_r)_h\right) =$$

$$= \frac{\partial}{\partial t}\left(\int_0^h v_z\, rdr\right) + \frac{\partial}{\partial z}\int_0^h v_z^2 rdr. \tag{A5}$$

In Eq. (A5), we have taken into account Eq. (7). As one can see the *LHS* of Eq. (10), Eq. (A5), has a divergent form.

Let us present the *RHS* of Eq. (10) as $RHS = A + B + C$, where

$$A = -\frac{1}{\rho}\int_0^h \frac{\partial P}{\partial z}rdr = -\frac{1}{\rho}\frac{\partial}{\partial z}\int_0^h Prdr + \frac{1}{\rho}P_h h\frac{\partial h}{\partial z} = -\frac{1}{\rho}\frac{\partial}{\partial z}\int_0^h Prdr +$$

$$h\frac{\partial h}{\partial z}\left(\frac{\gamma}{\rho}\frac{1+\left(\frac{\partial h}{\partial z}\right)^2 - h\frac{\partial^2 h}{\partial z^2}}{h\left(1+\left(\frac{\partial h}{\partial z}\right)^2\right)^{1.5}} - \frac{2\mu}{\rho}\frac{1}{\left(\frac{\partial h}{\partial z}\right)^2+1}\left(\left(\frac{\partial v_r}{\partial r}\right)_h - \left(\frac{\partial v_z}{\partial r} + \frac{\partial v_r}{\partial z}\right)_h\frac{\partial h}{\partial z} + \left(\frac{\partial v_z}{\partial z}\right)_h\left(\frac{\partial h}{\partial z}\right)^2\right)\right) =$$

$$= -\frac{1}{\rho}\frac{\partial}{\partial z}\int_0^h Prdr + \frac{\gamma}{\rho}\frac{\partial}{\partial z}\left(\frac{h}{\left(1+\left(\frac{\partial h}{\partial z}\right)^2\right)^{0.5}}\right) -$$

$$\frac{2\mu}{\rho}\frac{h\frac{\partial h}{\partial z}}{\left(\frac{\partial h}{\partial z}\right)^2+1}\left(\left(\frac{\partial v_r}{\partial r}\right)_h - \left(\frac{\partial v_z}{\partial r} + \frac{\partial v_r}{\partial z}\right)_h\frac{\partial h}{\partial z} + \left(\frac{\partial v_z}{\partial z}\right)_h\left(\frac{\partial h}{\partial z}\right)^2\right), \tag{A6}$$

$$B = \frac{\mu}{\rho}\int_0^h \frac{1}{r}\frac{\partial}{\partial r}\left(r\frac{\partial v_z}{\partial r}\right)rdr = \frac{\mu}{\rho}h\left(\frac{\partial v_z}{\partial r}\right)_h, \tag{A7}$$

$$C = \frac{\mu}{\rho}\int_0^h \frac{\partial^2 v_z}{\partial z^2}rdr = \frac{\mu}{\rho}\left(\frac{\partial}{\partial z}\int_0^h \frac{\partial v_z}{\partial z}rdr - \frac{\partial h}{\partial z}h\left(\frac{\partial v_z}{\partial z}\right)_h\right) =$$

$$= \frac{\mu}{\rho}\left(-\frac{\partial}{\partial z}\int_0^h \left(\frac{1}{r}\frac{\partial}{\partial r}(rv_r)\right)rdr - \frac{\partial h}{\partial z}h\left(\frac{\partial v_z}{\partial z}\right)_h\right) =$$

$$= \frac{\mu}{\rho}\left(-\frac{\partial}{\partial z}(h(v_r)_h) - \frac{\partial h}{\partial z}h\left(\frac{\partial v_z}{\partial z}\right)_h\right). \tag{A8}$$

In these equations we have used Eqs. (A3), (A4), and (3), and substituted $P_h$ from Eq. (9) into Eq. (A6).

Collecting all terms of Eqs. (A6) - (A8) we obtain

$$RHS = A + B + C = \frac{\partial}{\partial z}\left(\frac{\gamma}{\rho}\frac{h}{\left(1+\left(\frac{\partial h}{\partial z}\right)^2\right)^{0.5}} - \frac{1}{\rho}\int_0^h P r dr - \frac{\mu}{\rho}h(v_r)_h\right) + D, \tag{A9}$$

where

$$D = \frac{\mu}{\rho}\frac{h}{\left(\frac{\partial h}{\partial z}\right)^2+1}\left(2\frac{\partial h}{\partial z}\left(\frac{\partial v_r}{\partial r}\right)_h + \left(\frac{\partial v_z}{\partial r}\right)_h\left(1-\left(\frac{\partial h}{\partial z}\right)^2\right) - 2\left(\frac{\partial v_r}{\partial z}\right)_h\left(\frac{\partial h}{\partial z}\right)^2 - \left(\frac{\partial v_z}{\partial z}\right)_h\frac{\partial h}{\partial z}\left(1-\left(\frac{\partial h}{\partial z}\right)^2\right)\right) \tag{A10}$$

is the non-divergent part of $RHS$. Let us present $D$ in a divergent form. For this, in the brackets of Eq. (A10), we add and subtract $-2\frac{\partial h}{\partial z}\frac{\partial v_z}{\partial z}$ and $\left(\frac{\partial v_r}{\partial z}\right)_h$ that reduce Eq. (A10) to the following form

$$D = \frac{\mu}{\rho}\frac{h}{\left(\frac{\partial h}{\partial z}\right)^2+1}\left(2\frac{\partial h}{\partial z}\left(\frac{\partial v_r}{\partial r} - \frac{\partial v_z}{\partial z}\right)_h + \left(1-\left(\frac{\partial h}{\partial z}\right)^2\right)\left(\frac{\partial v_r}{\partial z} + \frac{\partial v_z}{\partial r}\right)_h + \left(\left(\frac{\partial v_z}{\partial z}\right)_h\frac{\partial h}{\partial z} - \left(\frac{\partial v_r}{\partial z}\right)_h\right)\left(\left(\frac{\partial h}{\partial z}\right)^2+1\right)\right) =$$

$$= \frac{\mu}{\rho}\left(h\left(\frac{\partial v_z}{\partial z}\right)_h\frac{\partial h}{\partial z} - h\left(\frac{\partial v_r}{\partial z}\right)_h\right) = \frac{\mu}{\rho}\left(-\left(\frac{\partial(v_r r)}{\partial r}\right)_h\frac{\partial h}{\partial z} - h\left(\frac{\partial v_r}{\partial z}\right)_h\right) =$$

$$= -\frac{\mu}{\rho}\left(h\frac{\partial h}{\partial z}\left(\frac{\partial v_r}{\partial r}\right)_h + \frac{\partial h}{\partial z}(v_r)_h + h\left(\frac{\partial v_r}{\partial z}\right)_h\right) = -\frac{\mu}{\rho}\frac{\partial}{\partial z}((v_r)_h h). \tag{A11}$$

In Eq. (A11), we have taken into account Eq. (8) and then used Eq. (3). Combining Eqs. (A5), (A9) and (A11) we present Eq. (10) in the following divergent form

$$\frac{\partial}{\partial t}\left(\int_0^h v_z r dr\right) + \frac{\partial}{\partial z}\left(\int_0^h \left(v_z^2 + \frac{P}{\rho}\right)r dr - \frac{\gamma}{\rho}\frac{h}{\left(1+\left(\frac{\partial h}{\partial z}\right)^2\right)^{0.5}} + \frac{2\mu}{\rho}(v_r)_h h\right) = 0. \tag{A12}$$

**Appendix B: Derivation of Eq. (21) from the slender jet model [1]**

The set of the slender jet equations [1] can be presented in the following form:

$$\frac{\partial v_z^0}{\partial t} + v_z^0\frac{\partial v_z^0}{\partial z} = -\frac{\gamma}{\rho}\frac{\partial}{\partial z}\left(\frac{1}{R_1} + \frac{1}{R_2}\right) + \frac{3\mu}{\rho}\frac{1}{y}\frac{\partial}{\partial z}\left(y\frac{\partial v_z^0}{\partial z}\right), \tag{B1}$$

$$\frac{\partial y}{\partial t} + \frac{\partial}{\partial z}(y v_z^0) = 0, \tag{B2}$$

$$\frac{1}{R_1} + \frac{1}{R_2} = \frac{y+\frac{1}{2}\left(\frac{\partial y}{\partial z}\right)^2 - \frac{y}{2}\frac{\partial^2 y}{\partial z^2}}{\left(y+\frac{1}{4}\left(\frac{\partial y}{\partial z}\right)^2\right)^{1.5}}. \tag{B3}$$

Multiplying Eq. (B1) by $y$ and then using Eq. (B2), after simple algebra, we obtain

$$\frac{\partial}{\partial t}(v_z^0 y) + \frac{\partial}{\partial z}\left(y(v_z^0)^2 + \frac{\gamma}{\rho} y \frac{y + \frac{1}{2}\left(\frac{\partial y}{\partial z}\right)^2 - \frac{y}{2}\frac{\partial^2 y}{\partial z^2}}{\left(y + \frac{1}{4}\left(\frac{\partial y}{\partial z}\right)^2\right)^{1.5}} - \frac{3\mu}{\rho} y \frac{\partial v_z^0}{\partial z}\right) - \frac{\gamma}{\rho} \frac{y + \frac{1}{2}\left(\frac{\partial y}{\partial z}\right)^2 - \frac{y}{2}\frac{\partial^2 y}{\partial z^2}}{\left(y + \frac{1}{4}\left(\frac{\partial y}{\partial z}\right)^2\right)^{1.5}} \frac{\partial y}{\partial z} = 0, \quad (B4)$$

Direct calculation shows that the last term in the LHS of Eq. (B4) can be presented in the following form,

$$\frac{\gamma}{\rho} \frac{y + \frac{1}{2}\left(\frac{\partial y}{\partial z}\right)^2 - \frac{y}{2}\frac{\partial^2 y}{\partial z^2}}{\left(y + \frac{1}{4}\left(\frac{\partial y}{\partial z}\right)^2\right)^{1.5}} \frac{\partial y}{\partial z} = \frac{\gamma}{\rho} \frac{\partial}{\partial z}\left(\frac{2y}{\left(y + \frac{1}{4}\left(\frac{\partial y}{\partial z}\right)^2\right)^{0.5}}\right), \quad (B5)$$

Substituting Eq. (B5) into Eq. (B4) yields

$$\frac{\partial}{\partial t}(v_z^0 y) + \frac{\partial}{\partial z}\left(y(v_z^0)^2 - \frac{3\mu}{\rho} y \frac{\partial v_z^0}{\partial z} - \frac{\gamma}{\rho} y^2 \frac{1 + \frac{1}{2}\frac{\partial^2 y}{\partial z^2}}{\left(y + \frac{1}{4}\left(\frac{\partial y}{\partial z}\right)^2\right)^{1.5}}\right) = 0. \quad (B6)$$